\documentclass[a4paper,UKenglish]{lipics}
\usepackage{xspace} 
\usepackage{proof}
\usepackage{booktabs}	   		% Pacote para deixar tabelas mais bonitas.
\usepackage{qtree}			% Tableaux 
\usepackage{microtype}%if unwanted, comment out or use option "draft"

\title{\TL tutorial: An approach to learning Logic by proving and refuting}
\titlerunning{\TL tutorial: proving and refuting} %optional, in case that the title is too long; the running title should fit into the top page column

\author[1]{Patrick Terrematte}
\author[2]{Jo\~ao Marcos}
\affil[1]{Federal University of Rio Grande do Norte (UFRN)\\  Department of Informatics and Applied Mathematics (DIMAp)\\
Group for Logic, Language, Information, Theory and Applications (LoLITA) \\  Natal -- RN -- Brazil\\
  \texttt{jmarcos@dimap.ufrn.br}}
\affil[2]{Federal University of Rio Grande do Norte (UFRN)\\  Digital Metropolis Institute (IMD)\\
Group for Logic, Language, Information, Theory and Applications (LoLITA) \\  Natal -- RN -- Brazil\\
  \texttt{patrickt@imd.ufrn.br}}
\authorrunning{P. Terrematte and J. Marcos} %mandatory. First: Use abbreviated first/middle names. Second (only in severe cases): Use first author plus 'et. al.'

\Copyright{Patrick Terrematte and Jo\~ao Marcos}%mandatory, please use full first names. LIPIcs license is "CC-BY";  http://creativecommons.org/licenses/by/3.0/

\subjclass{K.3.1 Computer Uses in Education; F.4.1 Mathematical Logic}% mandatory: Please choose ACM 1998 classifications from http://www.acm.org/about/class/ccs98-html . E.g., cite as "F.1.1 Models of Computation". 
\keywords{Teaching Logic; Didactic Software; Proof Theory; Refutations.}% mandatory: Please provide 1-5 keywords
\serieslogo{logo_ttl}%please provide filename (without suffix)
\volumeinfo%(easychair interface)
  {M. Antonia {Huertas}, Jo\~ao {Marcos}, Mar\'ia {Manzano}, Sophie {Pinchinat}, \\
  Fran\c{c}ois {Schwarzentruber}}% editors
  {5}% number of editors: 1, 2, ....
  {4th International Conference on Tools for Teaching Logic}% event
  {1}% volume
  {1}% issue
  {233}% starting page number
\EventShortName{TTL2015}
\newcommand{\bo}{\textbf}

\newcommand{\e}{\emph}
\newcommand{\ep}{\emph}

\renewcommand{\t}{\textsc}

\renewcommand{\tt}{\texttt}
\newcommand{\noi}{\noindent}

\newcommand{\TL}{\t{TryLogic}\xspace}

\newcommand{\A}{\ensuremath{\alpha}\xspace}   
\newcommand{\B}{\ensuremath{\beta}\xspace}
\newcommand{\C}{\ensuremath{\gamma}\xspace}

\newcommand{\Aa}{\textsf{p}}   \newcommand{\Ab}{\textsf{q}}   \newcommand{\Ac}{\textsf{r}}
\newcommand{\Ad}{\textsf{s}}   \newcommand{\Ae}{\textsf{t}}   \newcommand{\Af}{\textsf{u}}

\newcommand{\bii}{\ensuremath{\leftrightarrow}}

\begin{document}

\maketitle

\begin{abstract}
Aiming to offer a framework for blended learning to the teaching of proof theory, the present paper describes an interactive tutorial, called \TL, teaching how to solve logical conjectures either by proofs or refutations. The paper also describes the integration of our infrastructure with the Virtual Learning Environment \tt{Moodle} through the IMS Learning Tools Interoperability specification, and evaluates the tool we have developed.\\
 \end{abstract}

\section{Introduction}
\label{TryS}

\begin{flushright}
\begin{minipage}{.7\textwidth}
\scriptsize
\e{``(...) Knowledge never hurts --- what hurts is helplessness, the futility of banging your head against a brick wall without finding either proof or disproof. I have often spent weeks trying to prove a false statement --- and when I learned that it is false, I felt victorious. Progress was made, knowledge was acquired, one more step toward the truth was taken.'' }

\t{--- Paul R. Halmos} \\
\ep{I Want to be a Mathematician: An Automathography} (1985) p.~91.\\
\end{minipage}
\end{flushright}

\noi Logic permeates computing and provides essential tools for dealing with data structures. 
The curricula of discrete mathematics and logic courses emphasize a kind of verificationist approach through the teaching of techniques that focus on proving, instead of approaches that encourage also disproving, or refuting by way of counter-examples. 
The main afteraffect of such approach is a common misunderstanding by the students of the boundaries between a deductive framework and a semantic refutation. 

Logic is essential, in particular, for verifying correctness of code.  In checking that an iteration behaves as expected, for instance, one has to: (i)~carefully choose an appropriate boolean condition to test, (ii)~check whether the given initial conditions of an iteration imply the required postcondition, and usually also (iii)~prove that the execution of the corresponding code terminates. 
In case the implication mentioned in step~(ii) is refuted, then the iteration code does not implement the desired specifications for the postcondition. Conversely, if it is shown that the preconditions do imply the postcondition, then the iteration does satisfy the desired specification. The heuristic used in the analysis of this sort of situation involves basically the challenge of finding out whether certain conjectures can be proved or refuted. 
The fact that the two latter tasks, proving and refuting, are complementary is a very practical application of the soundness and completeness metatheorems.

The aim of the present paper is to present a tool for teaching the use of logical reasoning to verify conjectures for which it has not previously been determined whether they are provable or refutable. 
{One of the main goals of our tool is to teach how to construct a fully justified counter-example to witness the falsity of a given conjecture.} 
This tool implements some of the teaching principles discussed in~\cite{jmarcos:TTL2015}.

In the current state-of-the-art, an approach not unlike ours is used in Bornat's~\cite{bornat2005}, where Logic (formal deductive proof, formal semantic disproof and program specification) is presented with the help of the proof assistant \tt{J$\forall$p$\exists$}. 
{Many other existing tools also combine the teaching of Proof Theory with Formal Semantics}, e.g.\ \tt{AproS Project}, \tt{Panda}, \tt{Tarski's World}, \tt{Fitch} and \tt{Boole}. 
On top of those methodologies and tools, our contribution is to track learning beyond the mere use of proof strategies. 
Continuing the work presented by Terrematte \ep{et al.} \cite{Terrematte2011}, here we present an interactive tutorial to guide the student through the process of learning by trial and error: the \TL\footnote{Available  at \url{http://lolita.dimap.ufrn.br/trylogic}.}.

\section{The \TL tutorial for proving and refuting}
 
\begin{flushright}
\begin{minipage}{.7\textwidth}
\scriptsize
\e{``(...)  mathematics is not a deductive science. When you try to prove a theorem, you do not list the hypotheses, and then start to reason. What you do is trial and error, experimentation, guesswork. You want to find out what the facts are, and what you do is in that respect similar to what a laboratory technician does, but it is different in its degree of precision and information.''}

\t{--- Paul R. Halmos} \\
\ep{I Want to be a Mathematician: An Automathography} (1985) p.~321.\\
\end{minipage}
\end{flushright}

\noi 
Logic courses represent a pedagogical challenge and the re\-cor\-ded number of failures and discontinuities in them is often high.  
On that track, the main goal of \TL is to diminish the cognitive overload through a step-by-step tutorial, presenting different topics of logic related to the process of proving or refuting logical conjectures. Our tool \TL aims to:

\begin{itemize}
  \item present a set of lessons in Propositional Logic that exemplify the task of proving in natural deduction (theory~$N_p$) and refuting in a formal semantics (theory \tt{Sem$_p$}) through \tt{Coq}; 
  \item organize Logic in an interactive way and provide self-evaluation tasks to students;
  \item provide the teacher with a follow-up activity report on the lessons completed by the student at \tt{ProofWeb}, and provide a multi-language infrastructure for human-machine interaction.
\end{itemize}

The framework is implemented by combining the following tools:
 
\begin{itemize}
  \item  \tt{ProofWeb}\footnote{Available  at \url{http://prover.cs.ru.nl/}.}: an open source software for teaching Natural Deduction which provides interaction between some proof assistants (\tt{Coq}\footnote{Available  at \url{https://coq.inria.fr/}},\tt{Isabelle}, \tt{Lego}) and a Web interface~\cite{hendriks2010}.
  \item \tt{Conjectures Generator}\footnote{Available at \url{http://lolita.dimap.ufrn.br/logicamente-ge/} and it is open source code is available at \url{http://github.com/terrematte/logic-propgenerator}.}: a tool for task generation of a set of conjectural arguments (i.e.\ a set of premises with a formula that may follow or not from these premises). This tool was developed by our group.
  \item  \tt{TryOCaml}\footnote{Available at \url{http://try.ocamlpro.com/}}: an infrastructure consisting of an interactive tutorial for teaching and interaction with the functional programming language OCaml. 
  \item \tt{Moodle}: a well-known Virtual Learning Environment (VLE) that helps in organizing contents and educational
activities.
  \item  \tt{IMS Basic Learning Tools Interoperability}\footnote{The IMS LTI was developed in 2006 by IMS Global Learning Consortium and is available at \url{http://www.imsglobal.org/lti/index.html}.} (IMS LTI): a specification for the implementation and integration of educational tools.
\end{itemize}

\begin{figure}[htpb]
\centering
\caption{ Lessons on \TL integrated to \tt{Moodle}}
\includegraphics[width=14cm]{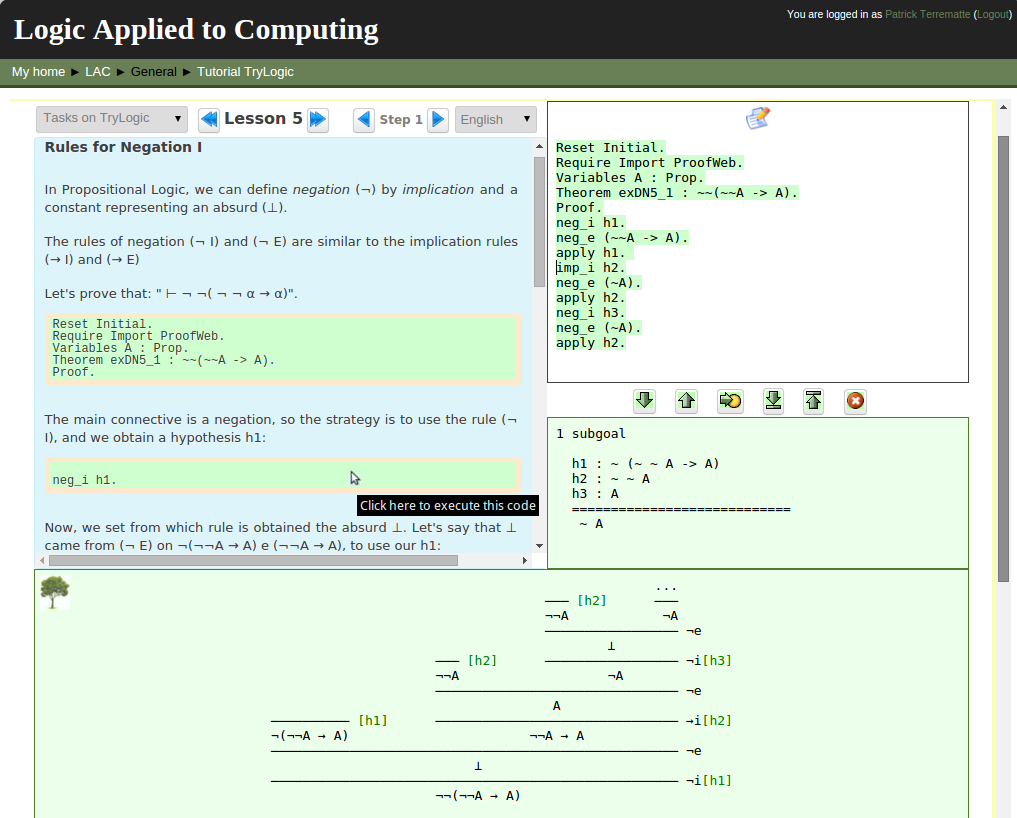}
\label{fig:trylogic}
\end{figure}

The \TL was developed as branch of the \tt{TryOCaml} project, i.e., all lessons and interaction follow the same architecture of the latter. We implemented the \tt{Sem$_p$} theory in \tt{Coq} and integrated the whole system with \tt{ProofWeb} and \tt{Moodle}.  With a goal of centralizing the use of our tools, we have used the specification IMS LTI, which is, according to the survey \cite{Alario-Hoyos2010}, one of the most representative alternatives to infrastructure integration between teaching platforms. Using this specification, any collaborator who wishes to use \TL in any VLE can add it as an external tool. This way, it is not necessary to install \TL nor is it necessary to obtain special access permission for the server in which it is being installed.

\subsection{The Conjectures Generator}

\noi The \tt{Conjectures Generator} for Propositional Logic was implemented through a formula generator with the SAT-solver \tt{Limboole}\footnote{ Available at \url{http://fmv.jku.at/limboole/}.} to evaluate propositional formulae.

The \tt{Conjectures Generator} creates conjectures in the format of individualized tasks for \tt{Coq}, directly in each student's area in \tt{ProofWeb}. The students receive each task in a template for them to try and prove (showing that $\Gamma \vdash \A$) in the~$N_p$ theory and another one for them to try and refute (showing that $\Gamma \nvDash \A$) in \tt{Sem$_p$} theory.  Of course, soundness and completeness connecting the two theories guarantee that only one of these tasks can be fulfilled.

The \tt{Conjectures Generator} was implemented with requirements that allow one to establish some connections of relevance between the premises and the conclusion, namely: that both the conclusion and the conjunction of the premises should be contingent, and that each formula of the premises must share at least one atom with the conclusion. Other settings are available, e.g.\ choosing the number of conjectures, number of premises, number of distinct atoms, selecting the connectives and a range for the complexity of the formulae; also, the user may decide if in the generated conjectures all premises are necessary to prove the conclusion, and if the collection of generated conjectures are all provable, all refutable, or are evenly divided into provable and refutable, to be randomly assigned to the students.

Through the available settings, the \tt{Conjectures Generator} is a useful tool for the teacher who wishes to evaluate propositional arguments through truth-tables, tableaux, natural deduction, resolution methods and even produce tasks concerning the evaluation of arguments.

\section{Propositional logic for proving and refuting} 
 
\begin{flushright}
\begin{minipage}{.7\textwidth}
\scriptsize
\e{``(...)  Every genuine test of a theory is an attempt to falsify it, or refute it.'' }

\t{--- Karl Raimund Popper} \\
\ep{Conjectures and Refutations: The Growth of Scientific Knowledge} (1963) p.~36.
\end{minipage}
\end{flushright}

\noi To \bo{prove} a conjecture $\Gamma \vdash \A$ in propositional logic with Natural Deduction it is necessary to build a derivation tree to witness it.  Our students were taught to do this using the rules of a theory we call~$N_p$, which is essentially the same that is natively implemented on \tt{ProofWeb}\footnote{Check the \tt{ProofWeb}'s manual at \url{https://prover.cs.ru.nl/man.pdf}}, following the usual style of natural deduction introduced by Gerhard Gentzen in 1935. 
As an illustration, the rules for disjunction  are the following:
$$
\infer[^{(\vee I_0)}]{\A \vee \B}{\B}
\qquad
\infer[^{(\vee I_1)}]{\A \vee \B}{\A}
\qquad
\infer[^{(\vee E):m,n}]{\C}{\A\vee \B & \infer*{\C}{\infer[^{m}]{\A}{}} & \infer*{\C}{\infer[^{n}]{\B}{}}}
$$ 

In constrast, in an approach involving formal semantics, we build a refutation tree by using the notions of valuation and satisfaction. A valuation~$v$ maps formulae to truth-values. An argument is refutable ($\Gamma \nvDash \A$) if there is a valuation~$v$ that satisfies all formulae in~$\Gamma$ and simultaneously falsifies~$\A$ ($v \Vdash \Gamma$ and $v \nVdash \A$). To \bo{refute} conjectures the students need to build refutation trees on the  \tt{Sem$_p$} theory. 
The rules of \tt{Sem$_p$} compositionally manipulate satisfaction of formulae by a valuation~$v$, and they are the following:

$$
\infer[^{(\neg~T)}]{v \Vdash \neg \A}{v \nVdash \A }
\qquad
\infer[^{(\neg~F)}]{v \nVdash \neg \A}{v \Vdash \A}
\qquad
\infer[^{(\top)}]{v \Vdash \top}{}
\qquad
\infer[^{(\bot)}]{v \nVdash \bot}{}
$$
$$
\infer[^{(\land~T)}]{v \Vdash \A \land \B}{v \Vdash \A & v \Vdash \B}
\qquad
\infer[^{(\land~F1)}]{v \nVdash \A \land \B}{v \nVdash \A }
\qquad
\infer[^{(\land~F2)}]{v \nVdash \A \land \B}{v \nVdash \B}
$$
$$
\infer[^{(\vee~T1)}]{v \Vdash \A \vee \B}{v \Vdash \A }
\qquad
\infer[^{(\vee~T2)}]{v \Vdash \A \vee \B}{v \Vdash \B}
\qquad
\infer[^{(\vee~F)}]{v \nVdash \A \vee \B}{v \nVdash \A & v \nVdash \B}
$$
$$
\infer[^{(\to~T1)}]{v \Vdash \A \to \B}{v \nVdash \A}
\qquad
\infer[^{(\to~T2)}]{v \Vdash \A \to \B}{v \Vdash \B }
\qquad
\infer[^{(\to~F)}]{v \nVdash \A \to \B}{v \Vdash \A & v \nVdash \B}
$$
$$ 
\infer[^{(\bii~T1)}]{v \Vdash \A \bii \B}{v \Vdash \A & v \Vdash \B}
\qquad
\infer[^{(\bii~T1)}]{v \Vdash \A \bii \B}{v \nVdash \A  & v \nVdash \B}
$$
$$
\infer[^{(\bii~F1)}]{v \nVdash \A \bii \B}{v \Vdash \A & v \nVdash \B}
\qquad
\infer[^{(\bii~F2)}]{v \nVdash \A \bii \B}{v \nVdash \A & v \Vdash \B}
$$

\noi
With these rules, the \tt{Sem$_p$} theory allows us to show that a given sentence is not a semantic consequence of a given set of premises. Here is a full example of a refutation tree: 
$$
\infer=[^{(by \vDash)}]{ \Ab \vee (\Ac \to \Aa) , (\neg \Aa \vee \neg \Ac) \to \Ab \nvDash \neg(\Aa \bii \Ab) \to (\neg \Ac \land \Aa)}{
  \infer[^{(\vee T2)}]{v \Vdash \Ab \vee  (\Ac \to \Aa)}{
    	\infer[^{(\to T2)}]{v \Vdash \Ac \to \Aa}{
	  \infer[^{}]{v \Vdash \Aa }{}
	}
  }
  &
  \infer[^{(\to T1)}]{v \Vdash (\neg\Aa \vee \neg \Ac) \to \Ab }{
    \infer[^{(\vee F)}]{v \nVdash \neg\Aa \vee \neg \Ac}{
      \infer[^{(\neg F)}]{v \nVdash \neg\Aa}{
	\infer[^{}]{v \Vdash \Aa }{}
      }
      &    
      \infer[^{(\neg F)}]{v \nVdash \neg\Ac}{
	\infer[^{}]{v \Vdash \Ac }{}
      }
    }
  }
  &
  \infer[^{(\to F)}]{v \nVdash \neg(\Aa \bii \Ab) \to (\neg \Ac \land \Aa)}{
    \infer[^{(\neg T)}]{v \Vdash \neg(\Aa \bii \Ab)}{
	\infer[^{(\bii F)}]{v \nVdash \Aa \bii \Ab}{
	   \infer[^{}]{v \Vdash \Aa }{}
	   &
	   \infer[^{}]{v \nVdash \Ab }{}
	}}
	&
	\infer[^{(\land F1)}]{v \nVdash \neg \Ac \land \Aa}{
	   \infer[^{(\neg F)}]{v \nVdash \neg \Ac }{
	   \infer[^{}]{v \Vdash \Ac }{}
	   }
	}
  }}
$$
In a bottom-up reading each connective in  \tt{Sem$_p$} has rules that provide a sufficient condition for a valuation $v$ to satisfy (or falsify) a given sentence. On the other hand, in a top-down reading, the application of the rules represent a semantic inference. In the branches of the refutation tree one finds statements in the form $v \Vdash \A$ or $v \nVdash \A$. In the leaves, one finds statements such as $v \Vdash \Aa$ or $v \nVdash \Aa$, where $\Aa$ is an atomic formula.  A refutation tree represents thus a fully justified counter-model to a given conjecture.
Note that the rules are analytical, i.e.\ the premises of each rule contains statements over subformulas of the formula in the conclusion of the rule. This ultimately means that the leaves of an exhausted tree are always over atomic formulae.

The general backward strategy for building a refutation tree follows these steps:

\begin{enumerate}
 \item Assume that the conjecture is refutable, i.e.\ that there is a valuation $v$  that satisfies its premises and falsifies its purported conclusion.
 \item Exhaustively explore and justify step-by-step with the \tt{Sem$_p$} theory the consequences of the initial assumption that the conjecture is refutable. The exhaustive exploration means it may be necessary to backtrack to try other possible rules.
 \item Check if the valuation consistently satisfies or falsifies each propositional atom involved, i.e.\ it cannot be that a valuation $v$ satisfies some atom $\Aa$ ($v \Vdash \Aa$) and falsifies the same atom $\Aa$ ($v \nVdash \Aa$) in the same refutation tree.
\end{enumerate}

Note that the above strategy only applies to refuting contingent or contradictory formulae, and to exhibiting witnesses to invalid arguments. If one does not manage to build a refutation tree after an exhaustive exploration of the possibilities, this means that there does not exist a valuation $v$ such $v \Vdash \Gamma$ and $v \nVdash \alpha$. Therefore, by the relation of semantic consequence we know that $\Gamma \vDash \alpha$.  Thus, from the completeness theorem ($\Gamma \vDash \alpha \Rightarrow  \Gamma \vdash \alpha$), one knows that the conjecture can be proved, i.e.\ that it is possible to build a derivation tree in the $N_p$ theory.

\subsection{Some logical and pedagogical remarks}

\begin{figure}[htbp]
\centering
\scriptsize
  \caption{\label{fig:taticas} \ep{Script} for refutation on \tt{ProofWeb}}
    \begin{tabular}{|p{5.5cm}|p{6.2cm}|}
\toprule
\hline
\vspace{-2.8cm}
\begin{verbatim}

Reset Initial. 
Require Import Semantics.
Parameter A B C D : Prop.
Hypothesis f1 : (v ||-/- A).
Hypothesis f2 : (v ||-/- B).
Theorem sem_ex2 :  
(v ||-/- ((~A->B) /\ (~A/\~B))).
Proof.  	
conjF1.
impF. 
negT.
apply f1.
apply f2.

\end{verbatim}
&
\includegraphics[width=6.2cm]{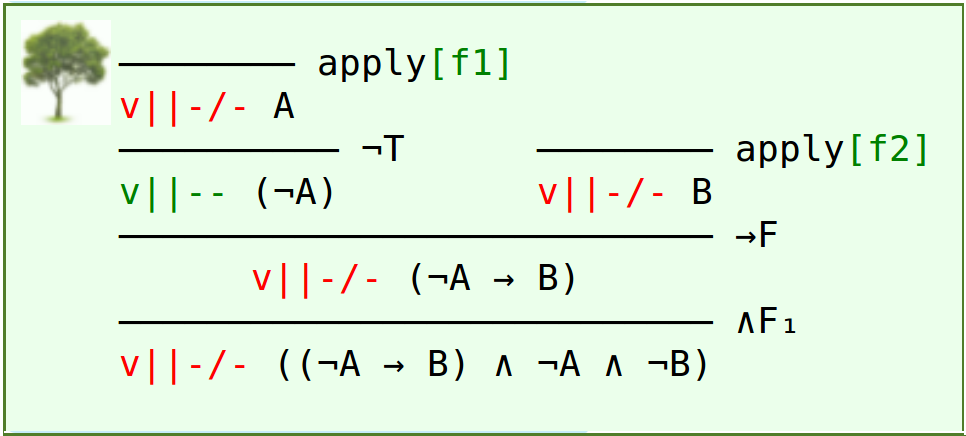}
\\ \hline
\footnotesize
a) \ep{Script of refuting tacticals}
&
\footnotesize
b) \ep{\label{fig:sem-sem-ex1} Resulting tree on \TL}/\tt{ProofWeb}
\\ \hline   
     \bottomrule 
    \end{tabular}
\end{figure}

Each connective rule of the \tt{Sem$_p$} theory is implemented by {tacticals} in \tt{Coq} and we extended the \tt{ProofWeb} system to display the corresponding refutation trees, as illustrated in Fig.~\ref{fig:taticas}. 

{One of the pedagogical advantages of \tt{ProofWeb} is its coherence with the \textit{modality effect} of Cognitive Load Theory \cite[p.129]{Sweller2011}. The idea is that two well connected sources of information reinforce the organizing process and facilitate the transfer of information to the long-term memory information store.} 
The multiple representation is applied in Fig.~\ref{fig:taticas}, where we can observe that without the proof script being inserted on the left-hand side (a) it would not be possible to check the object on the right-hand side (b) for the tactical sequence represents a justification for the refutation produced. 
The refutations are not static (b), but in fact, correspond to the dynamic linear process of their construction on side (a). Thus, the visualization of (b) has didactic value as it is also useful to the communicability of the refutation structure in (a).

\begin{figure}[t]
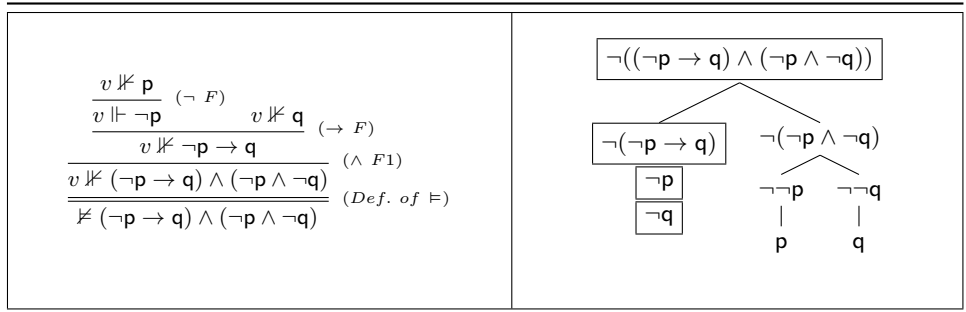

\centering
\footnotesize
  \caption{\label{fig:heu-tabl} Refutation tree \ep{versus} Tableaux Method}
    \begin{tabular}{|p{6.2cm}|p{5.5cm}|}
\toprule
\hline
\vspace{0.1cm}

$$
\footnotesize
\infer=[^{(Def.~of~\vDash)}]{\nvDash (\neg \Aa \to \Ab) \land (\neg\Aa \land \neg \Ab)}{
\infer[^{(\land~F1)}]{  v \nVdash (\neg \Aa \to \Ab) \land (\neg\Aa \land \neg \Ab) }{
  \infer[^{(\to~F)}]{v \nVdash \neg \Aa \to \Ab}{
    \infer[^{(\neg~F)}]{v \Vdash \neg \Aa}{v \nVdash \Aa}
    &
    v \nVdash \Ab
  }
}}
$$

&
\vspace{0.1cm}

\Tree
[.\framebox{$\neg ((\neg \Aa \to \Ab) \land (\neg\Aa \land \neg \Ab))$}
[.{\framebox{$\neg(\neg \Aa \to \Ab)$} \\
  \framebox{$\neg \Aa$} \\
   \framebox{$\neg \Ab$}} ]
[.{$\neg(\neg\Aa \land \neg \Ab)$}
[.{$\neg\neg\Aa$}
[.{$\Aa$} ]
]
[.{$\neg\neg\Ab$}
[.{$\Ab$} ]
]
]
]
\vspace{\baselineskip}

\\ \hline   
     \bottomrule 
    \end{tabular}
\end{figure}

The heuristic procedure for refutation presented here might be replaced by other deductive formalisms in Propositional Logic, such as the sequent calculus, resolution, tableaux or even truth-tables. However, we avoid the truth-table method for its fixed exponential computational cost ($2^n$, where $n$ is the number of distinct atoms in the conjecture) 
and its purely algorithmic character, which we judge not to have optimal pedagogical value. 
{Tableaux, on the other hand, are often very efficient in both tasks of proving and refuting}. However, they also make the procedure fully automatic. 
While this might be a desirable property from a computational viewpoint, from the didactic perspective we claim that tableaux create a conceptually undesirable overlap between deductive formalism and formal semantics. As a consequence of the exclusive use of tableaux, students are often led to build no appreciation at all for the distinction between Proof Theory and Formal Semantics.
To clarify the meaning of our semantic heuristics, check the comparison between the refutation tree and the tableau method in Fig.~\ref{fig:heu-tabl}:
Note in particular that in using \tt{Sem$_p$} the students are forced all the time to take decisions about which tableau branch they should want to explore.
It is known that in the worst-case scenarios tableaux might be much more costly than truth-tables \cite[p.62]{DAgostino1992}. A tableau is exhausted only when all the branches are fully explored, and this may depend on the ratio between the complexity of the formulae and the atoms that occur in them. Therefore, if a formula has higher complexity than the number of distinct propositional atoms, then the tableau analysis may be longer than the number of rows in the truth table. In contrast, our heuristic procedure is not fully automatic and the wise choice of which path to follow may introduce exponential speed-up. Ultimately, the use of \tt{Sem$_p$} simply requires the production of a sequence of formulae corresponding to an open branch of a tableau tree.

Our goal is to improve the logical intuition of students. Therefore, students are told that in cases where the semantic heuristics do not allow for a refutation, they should look for a derivation tree in $N_p$. On the other hand, when they are having trouble in proving, they might well try to refute the selected conjecture.

\subsection{Analysis of experiments with \TL}
 
\noi We performed several experiments in the years of 2012, 2013 and 2014 to evaluate \TL in blended learning use. The task of proving or refuting was given to {the average of 15 students per semester of Computer Science in the upper undergraduate course of \textit{Logic Applied to Computing}} at the Federal University of Rio Grande do Norte (DIMAp/UFRN), of which an average of 58\% have actively used the system. Through face-to-face classes we taught only using theoretical fundamentals, and our biggest challenge was teaching the computer-assisted task of proving or refuting exclusively through \TL. {The main task given to students was to prove or refute six to eight conjectures randomly assigned. These consisted in two conjectures per each of three or four levels of difficulty. For instance at first level (easiest), the conjectures have 3 distinct propositional atoms, with 3 premises and a complexity between 2 and~4 connectives per formula. The fourth and hardest level has 6 distinct propositional atoms, with~4 premises and a complexity of~4 to~6 of connectives.} {The learning goals are to practice formal proof and refutation heuristics, as well as to advance the understanding of soundness and completeness metatheorems. At the end of each experiment, students answered a questionnaire about their profile, their use of the available tools, their difficulties in solving the tasks and their theoretical understanding of the tasks.} 

Some general conclusions about the experiments are:
\begin{itemize}
 \item \TL provides the understanding of the deductive process in \tt{Coq} to students who had brief theoretical contact with the content of Natural Deduction.
 \item {The students consistently solved more refutable conjectures than provable ones, even if they have received in average an equal number of each kind of conjecture. For instance in Spring 2013, out of 60 solved conjectures, the students presented 43 refutations. It is possible to draw at least the following two interpretations for this phenomenon: that the search for a refutation tree is easier than the production of a natural deduction proof, or that the lessons for proving in Natural Deduction on \TL need to have an improved teaching strategy. The first interpretation is coherent with our learning goals, we aim to show that refutation is natural and necessary in Logic. As for the second interpretation, we feel it important to add that some conjectures given as task are really large and difficult to prove\footnote{For an example, a conjecture generated in the fourth level to be proved in Natural Deduction was this one:\\
 $\{\neg ( \Aa \vee ( ( \Ab \to ( \Af \bii \Ac ) )  \land ( \Ad \bii \Ac ) ) ),
\Ae \to ( \neg ( \Ac \vee ( \Ac \vee ( \Aa \bii \Aa ) ) ) ), 
 \neg ( ( \neg ( ( \Aa \bii ( \neg \Ae ) ) \bii \Af ) ) \vee \Aa ),
( ( ( ( \Ac  \land \Ad )  \land \Af ) \to ( \Ad \vee \Aa ) ) \bii \Af )  \land \Ad \} \vdash ( \Af \to ( \Af  \land ( \Ae \to ( \Ad \to \Ab ) ) ) ) \vee ( \Ac \bii \Aa ) $.} and this might explain the smaller number of produced formal proofs.}
 \item {A negative conclusion drawn from the questionnaire was that the practice involved on prove or refute, does not necessary imply the theoretical understanding of the metatheorems of completeness and soundness. }
 \item Using the theories $N_p$ and \tt{Sem}$_p$, implemented in \tt{Coq}, the student applies a heuristic for proving and refuting through justified and verified steps. This way, with \TL, the experimental process of `trial and error' is taught in a guided environment.
\end{itemize}

\section{Future Works and Final Remarks}
 
\begin{flushright}
\begin{minipage}{.81\textwidth}
\scriptsize
\e{``(...)  we teach mathematics to the engineers, physicists, biologists, psychologists, economists --- and mathematicians --- of the future. (...) It is not enough to teach them everything that's known---they must know also how to find out what has not yet been found.'' }

\t{--- Paul R. Halmos} \\
\ep{I Want to be a Mathematician: An Automathography} (1985) p.~322.
\end{minipage}
\end{flushright} 
 
\noi 
This paper presents an infrastructure of integrated tools for the teaching of Logic with focus on: (i)~an organized step-by-step presentation of the content of Natural Deduction and Propositional Semantics in a sequential and interactive way; (ii)~providing the student with interactive self-evaluation tasks; (iii)~the interaction with the \texttt{Conjectures Generator} and \TL with \tt{Moodle} through IMS LTI. {It is worth noting that, since the \TL is based on \tt{ProofWeb} and the lessons are structured on \tt{TryOCaml}, our infrastructure is extensible and customizable\footnote{The project can be forked from: \url{https://github.com/terrematte/trylogic}.} to build lessons on any other formal theory implemented on \tt{Coq} or \tt{Isabelle}, e.g.\  on Modal Logic, Number Theory, Set Theory, or Hoare Logic.} Our contributions to teaching Logic are part of an initiative that needs to be enhanced. Some opportunities for the extension of the project would include:

\begin{itemize}
   \item Producing lessons in English, Spanish, French and other languages.
   \item {Implementing in the \tt{Conjectures Gene\-ra\-tor} metrics of difficulty for derivations given by the size of normalized proofs and the number of uses of certain rules in the latter proofs.
   \item Extending the \tt{Conjectures Gene\-ra\-tor} and the theory \tt{Sem$_p$} to First Order Logic and producing new tasks and lessons of First Order Logic through \TL.}
\end{itemize}

\subparagraph*{Acknowledgements}
The authors acknowledge partial support by the Marie Curie project PIRSES-GA-2012-318986, funded by EU-FP7, and by CNPq~/ Brazil.
The authors also want to thank all undergraduate students of Computer Science and Computer Engineering who have contributed to the project during several semesters of the course of Logic Applied to Computing at DIMAp/UFRN. For the implementation of the \tt{Conjectures Generator}, a special acknowledgement should go to Elias Amaral.

%%
%% Bibliography
%%

%% Either use bibtex (recommended), but commented out in this sample

%\bibliography{dummybib}

%% .. or use bibitems explicitely

%\bibliography{abntex-references}

\begin{thebibliography}{}

\end{thebibliography}


\begin{thebibliography}{1}
\label{TryE}

\bibitem{Alario-Hoyos2010}
Carlos Alario-Hoyos and Scott Wilson.
\newblock Comparison of the main alternatives to the integration of external
  tools in different platforms.
\newblock In {\em Proceedings of the International Conference of Education,
  Research and Innovation, ICERI}, pages 3466--3476, 2010.

\bibitem{bornat2005}
Richard Bornat.
\newblock {\em Proof and Disproof in formal logic: an introduction for
  programmers}.
\newblock Oxford University Press, 2005.

\bibitem{DAgostino1992}
Marcello D'Agostino.
\newblock {\em Investigations into the complexity of some propositional
  calculi.}
\newblock PhD thesis, University of Oxford, 1992.

% \bibitem{VanDalen2004}
% Dirk~Van Dalen.
% \newblock {\em Logic and Structure}.
% \newblock Springer, 4th edition, 2004.

\bibitem{hendriks2010}
Maxim Hendriks, Cezary Kaliszyk, Femke Van Raamsdonk, and Freek Wiedijk.
\newblock {Teaching logic using a state-of-the-art proof-assistant}.
\newblock {\em Acta Didactica Napocensia}, 3(2):35--48, 2010.
%\newblock Available at:
%  \url{http://dppd.ubbcluj.ro/adn/article_3_2_4.pdf}.

\bibitem{jmarcos:TTL2015}
Jo{\~a}o Marcos.
\newblock Fail better: What formalized math can teach us about learning.
\newblock 4th International Conference on Tools for Teaching Logic, pages 119--128.

\bibitem{Sweller2011}
John Sweller, Paul Ayres, and S.~Kalyuga.
\newblock {\em Cognitive Load Theory}.
\newblock Explorations in the Learning Sciences, Instructional Systems and
  Performance Technologies. Springer, 2011.

\bibitem{Terrematte2011}
Patrick Terrematte, Fabr\'icio Costa, and Jo{\~a}o Marcos.
\newblock Logicamente: A {V}irtual {L}earning {E}nvironment for logic based on
  {L}earning {O}bjects.
\newblock In Patrick Blackburn, Hans Ditmarsch, Mar\'ia Manzano, and Fernando
  Soler-Toscano, editors, {\em Third International Conference Tools for
  Teaching Logic.}, volume 6680 of {\em Lecture Notes in Artificial
  Intelligence}, pages 223--230. Springer-Verlag, Berlin, 2011.

\end{thebibliography}

\newpage
\thispagestyle{empty}
{\ }

\end{document}